# A Rolling Blockchain
# for a Dynamic WSNs in a Smart City

Sergii Kushch, Senior Member, IEEE, Francisco Prieto-Castrillo

**Abstract**—Blockchain is one of the most popular topics for discussion now. However, most experts still see this technology as only part of Bitcoin, other crypto-currencies or money transfer systems. Often, new solutions, proposed by young researchers, are blocked by reviewers, only because these solutions can not be used for Bitcoins. However, Blockchain technology is more universal and can be used also in other areas, for example, in IoT, WSN and mobile devices.

This paper considers the implementation of Blockchain technology in sensor networks as an element of IoT. The concept of "Rolling Blockchain" was proposed, which can be used to build WSN with the participation of Smart Cars, as nodes of the network.

The order of block formation and structure in the chain is proposed and a mathematical model is created for it. We estimate the optimal number of WSN nodes, the number of connections between nodes, for specified network reliability values, was performed.

✦

## 1 INTRODUCTION

THE essence of Blockchain (BC) technology is the secure, distributed storage of any kind of information. BC can store data on transactions; on who, to whom and what amount of money has been transferred (cryptocurrencies, bank transactions). Currently, this is the main area in which Blockchain is used.

However, attempts are being made to apply it in other areas, for example, to record cargo during transportation, to manage "smart cities", create "smart contracts", and for the Internet-of-Things (IoT) and Wireless Sensor Networks (WSN) etc. [1], [2], [3], [4], [5], [6], [7], [8]. BC was conceived as a system that is completely protected from the substitution of information in existing blocks of the chain. This property makes us look for ways of using the BC technology as a method of protecting the information that is transmitted from various sensors and mobile devices. This also implies its storage, without any possibility of substituting part or all of the information. With respect to cryptocurrency, BC is the mainstay of Bitcoin's [9], [10] financial strength. It guarantees that information about money transfers between all the system participants is recorded during the entire period of the existence of the Bitcoin system.

BC is structured as a chain of blocks that contains information, consequently all the blocks of a chain are connected with each other. A block is filled with a group of records, and new blocks are always added to the end of the chain, apart of containing new information, new blocks duplicate the infor-mation contained in the previously created structural units of the system.

Construction of BC chains occurs on the basis of three main principles - distribution, openness and protection [11], [12]. Users of the system form a computer network. At the same time, each computer stores a copy of each of the blocks. This structure is provided by the interaction of "miners" who solve complex, expensive mathematical tasks. To solve them, it is necessary to spend both material resources (electricity, specialized "farms" for "mining"), and also the hardware capabilities for complex mathematical calculations known as Proof-of-Work (POW) [13]. The results of mining are collected in the BC and as the length of the chain increases with time, its reliability increases.

Also, with time, the complexity of the problem solved by the "miners" increases with the chain. All this requires an increase in both the computing power of "farms" and in the volume of devices that store the entire chain.

However, using BC on mobile devices, for example, in a smart sensor network, poses problem that makes using BC impossible; this is because the sensors do not have the computational resources to perform POW.

Another well-known problem is that of WSNs nodes, due to the limited volume of node batteries, have a limited period of operation and as a consequence the entire network is limited. In [14], [15], [16], [17], [18], it is shown that dependence on energy consumption is created by: the algorithm of operation (time of work, sleep, awake), the use of MAC-protocols, the amount of transmitted, received and processed information, data acquisition from sensors, and some other parameters. In the case of using POW, which is known as a very resource-intensive and energy-intensive task, the autonomous work of the nodes will be significantly reduced.

In addition, the standard structure of BC requires a complete connection between all elements of the network but this is not always possible and energetically advantageous.

————————————————

*S. Kushch, is with the Security and Trust research unit of the Bruno Kessler Foundation, via Sommarive, 18, Povo, 38123,TN, Italy
E-mail: kushch@yaros.co, skushch@fbk.es*
*Francisco Prieto-Castrillo with Media Laboratory, Massachusetts Institute of Technology, Cambridge, MA 02139-4307, USA; Harvard T.H. Chan School of Public Health, Harvard, Boston, MA 02115, USA; University of Salamanca, BISITE Research Group, Edificio I+D+i, 37008, Salamanca, Spain E-mail: fprieto@hsph.harvard.edu, fprieto@mit.edu, franciscop@usal.es.*

All these works provide key insights into the problem of network resilience, diffusion and consensus from different perspectives. But, according to the authors, a model of BC for devices with limited resources, for full and partially connected BC, is still missing. Therefore, in this paper we make an analysis of the conditions under which using BC in distributed sensor networks and IoT devices is possible. The question at issue is how to design BC without POW with a partial connectivity while maintaining robustness to failures and attacks. To this end, we developed several network models.

The paper is organized as follows. In Section 2 we formulate the problem. The results obtained in the study are presented in Section 3. Finally, we present the conclusions obtained from our research and discuss the possibilities for future work in Section 4.

## 2 Problem formulation

Blockchain can be conceived as chains of separated elements which are interconnected by hashing. There are three key factors in this process: a) which structure of blocks and chain is used, b) how the network is built, c) how consensus can be achieved. We elaborate on these elements below.

### 2.1 Network structure

The approach proposed in this work is to build a closed private BC network with unchanged complexity. The number of new blocks per minute is set by a constant value. The entire database is stored on the server. At the same time, the server is the node for the distributed P2P network of servers which uses a BC to account for the information received from the local sensor networks. This will ensure that the stored information remains unchanged, making its storage and recovery more secure in the event of an attack on a separate data server.

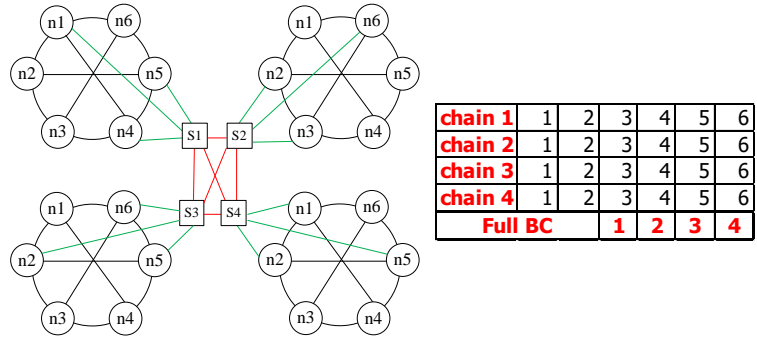

Fig. 1. A general network structure. The total network consists of local segments. Each segment has its own server for storing local BC. A distributed network of servers (which is marked-S1,..,S4) also stores a common BC consisting of local fragments.

An examples of such a networks can be: a power distribution company uses a network of smart meters and sensors for power consumption control, to account the use of electricity and monitor the status of the network; network that controls many medical devices (including portable), which are grouped into sub-clusters (eg within the same department of the hospital); WSN which monitors traffic on highway routes and consists of several segments etc. (Fig.1).

At the same time, these sensors and counters are combined into a local sensor network for each city. The servers of this organization in different cities will also be combined into a peer-to-peer network. It is in this network that a complete BC of all local sensor networks is stored. This will for example, avoid attacks on critical infrastructure by substituting information about the load in local networks, which can lead to power outages throughout the network.

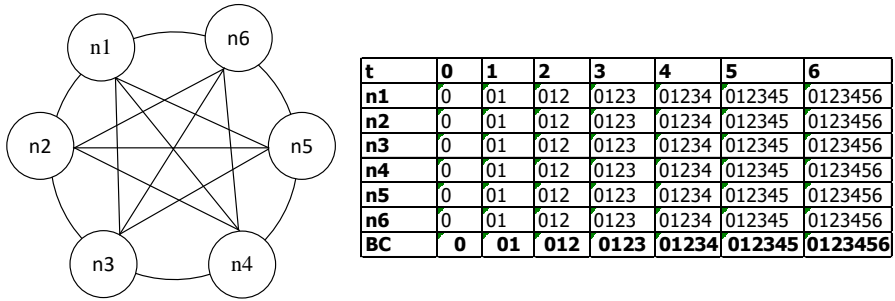

Fig. 2. Structure of the 6-element distributed network. The table shows the step-by-step formation of the chain. Blockchain is stored in each node of the chain.

Consider one segment of such a network consisting of six nodes. Fig.2 illustrates this example. In this case, each new block of information is forwarded to all nodes and BC is built in parallel on all the elements of the system.

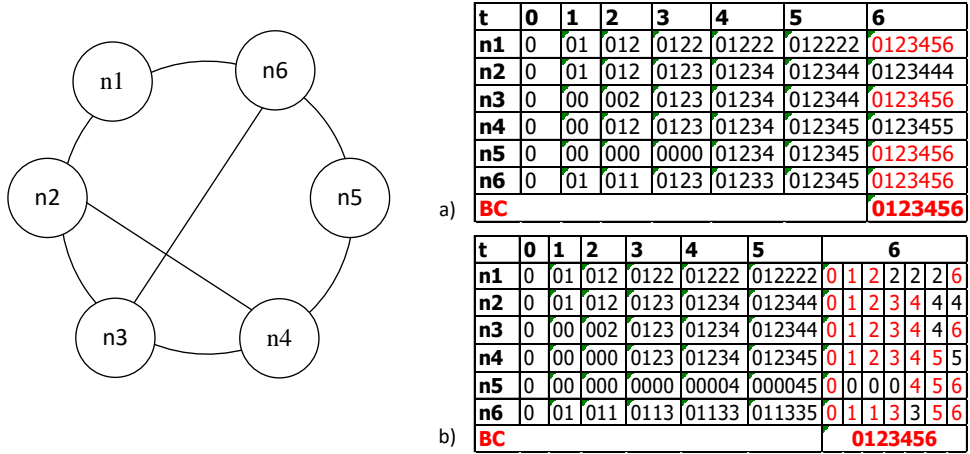

| t | 0 | 1 | 2 | 3 | 4 | 5 | 6 |
|---|---|---|---|---|---|---|---|
| n1 | 0 | 01 | 012 | 0122 | 01222 | 012222 | 0123456 |
| n2 | 0 | 01 | 012 | 0123 | 01234 | 012344 | 0123444 |
| n3 | 0 | 00 | 002 | 0123 | 01234 | 012344 | 0123456 |
| n4 | 0 | 00 | 012 | 0123 | 01234 | 012345 | 0123455 |
| n5 | 0 | 00 | 000 | 0000 | 01234 | 012345 | 0123456 |
| n6 | 0 | 01 | 011 | 0123 | 01233 | 012345 | 0123456 |
| BC | | | | | | | 0123456 |

a)

| t | 0 | 1 | 2 | 3 | 4 | 5 | 6 |
|---|---|---|---|---|---|---|---|
| n1 | 0 | 01 | 012 | 0122 | 01222 | 012222 | 0 1 2 2 2 2 6 |
| n2 | 0 | 01 | 012 | 0123 | 01234 | 012344 | 0 1 2 3 4 4 |
| n3 | 0 | 00 | 002 | 0123 | 01234 | 012344 | 0 1 2 3 4 4 6 |
| n4 | 0 | 00 | 000 | 0123 | 01234 | 012345 | 0 1 2 3 4 5 5 |
| n5 | 0 | 00 | 000 | 0000 | 00004 | 000045 | 0 0 0 0 4 5 6 |
| n6 | 0 | 01 | 011 | 0113 | 01133 | 011335 | 0 1 1 3 3 5 6 |
| BC | | | | | | | 0123456 |

b)

Fig. 3. The structure of a chain for the six iterations. During each iteration, information from the "active" node is sent to the "neighbors" which evaluate and give permission to write it to the end of the chain. The remaining nodes retain the previous state: a) a node sends a renewed BC to its neighbors, b) a node sends only one transaction which is added to the local BC of its neighbors.

It is considered that after the failure of one node the system becomes inoperative. However, in real conditions, the system can continue to work, but some of its parameters may change.

In these circumstances, we can say that the failure of one or more nodes leads to the transformation of a fully connected network to a partially connected network. Also, this situation is possible if the network was attacked and some connections or nodes are unavailable or compromised.

Consider the principle of chain formation. Each node contains a list that indicates: the list of nearest "neighbors", the node's activation time (sensor interrogation, formation of the BC block), and the order of sending the block to "neighbors". The use of such a list will solve at least two problems: it will optimize the time of sleep-awake-work, it will use the minimum-necessary power for sending-receiving information to and from neighbors. Also, it should be borne in mind that time of inclusion of each step differs from the time of inclusion of the remaining nodes and must satisfy the inequality: $T_{n1} < T_{n2} < \cdots < T_{nm}$.

The Blockchain is constructed as follows: step 0, all nodes record the genesis block 0, step 1- node 1 creates block 1 and send it to its neighbors. The inaccessible nodes repeat their previous state, etc. If a node is a failure node and sends a rejection to one of its neighbors, in this case it will be able to restore its state on the basis of consensus with its neighbors. As shown by tables a) and b) under Fig.3 - the rows represent nodes and the columns - the time points (iterations). The two BC building variants are possible.

First, the node that created the block, adds it to the BC in its own memory, and then sends the BC to its neighbors for validation. After the validation, the next step is to - send a confirmation that the BC is correct, after this confirmation the neighbors rewrite the renewed BC in their memory (the table a) of Fig.3.

Furthermore, after creating the block, the node sends only this block for validation. Then adds it to the end of the BC. The neighbors add one too (table b) in Fig.3. We see that the second method is more advantageous in terms of energy efficiency, as demonstrated above.

The process of BC validation is as follows: after forming the chain, the nodes produce an element-by-element verification of the final chain. If a block written in the chain of each node is confirmed (was written in) by more than 51% of the nodes at each iteration, such a block is written in the resultant BC. However, a possible situation is that there is a partial connection of nodes, and it is impossible to achieve 51% confirmation for some blocks. In this case, some blocks may be lost. The analysis of the number of lost blocks, depending on the number of connections in the distributed network, is presented by the authors in [19]

## 2.2 The structure of the chain

In this section, we discuss the structure of the chain that we propose to use for resource-constrained devices. The issues of constructing a distributed network structure were widely considered in [20], [21], [22]. In this work, we propose to use a chain consisting of several parts. Each part contains a limited number of generated blocks. Their number depends on the parameters and capabilities of the devices used in the IoT network. Below, we consider the proposed version of the block structure in more detail.

Notice, that the memory for storing BC is limited. Modern modules offered by manufacturers of IoT devices are usually limited from 1 to 8 MB of memory, most of which is used for storing the software that manages this device. Therefore, we are limited in the volume of BC (amount of blocks), which can be stored by each node.

The number of blocks stored by each node is limited by n, after which the n-1 blocks are removed from all nodes. Only the n block remains as the "zero" block (genesis block) for the next cycle. In this case, the full BC will be as follows(Fig.4):

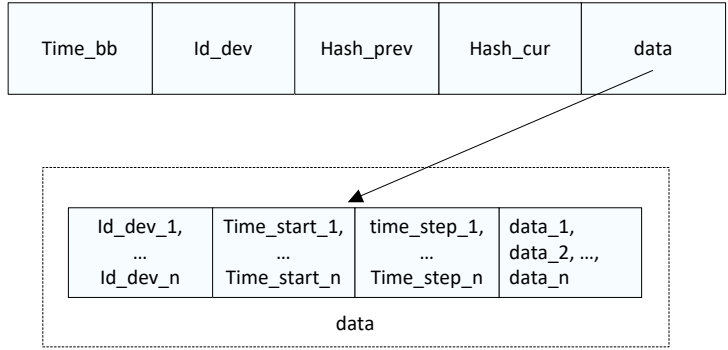

Fig. 4. The structure of Blockchain creating in the conditions of limited memory of WSN nodes.

The notation "Bn (B0) New cycle1" indicates that the block Bn is a zero block for the next cycle, etc. The number of blocks in the chain depends on the parameters of the "worst" memory device in the network, since the stored chain in each cycle should fit the available memory.

## 2.3 Block structure

Below, we consider the structure of each block in detail (Fig.5).

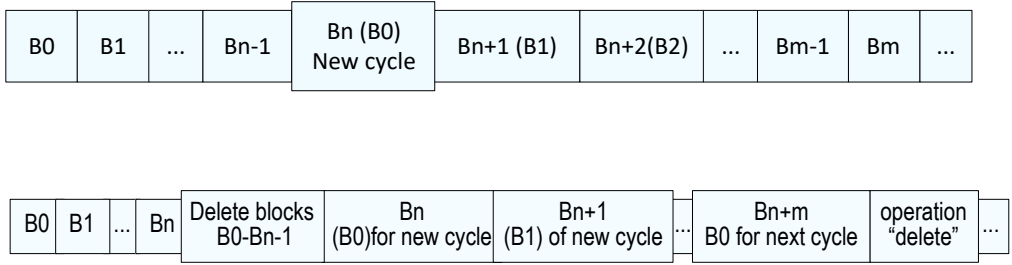

Fig. 5. General structure of the block. The top part of the figure shows the general block structure.. The bottom part shows the subblock "data" (the transactions block

It should be taken into account that each node can have several different sensors. Therefore, it is necessary to identify each sensor and the measurement time. If several measurements made over a period of time by each sensor (multi-segment transaction), are recorded in the block, a time stamp "step" must be added to show the time interval between measurements. In addition, given that the information transmitted by the sensors can be "closed", it is encrypted using cryptographic algorithms. This function is provided by modern IoT modules. In this case, the size of the Data block is set automatically, depending on the number of transactions.

## 2.4 Blockchain formation

There are two options for building this system:

1) When the node memory is "clearing", and only the last block remains (as B0 for the next cycle);

2) When only the first block is deleted from the stored chain, and a new block is recorded in the empty space that has been unallocated in the memory. For example, the block B0 is deleted and the block n+1 is recorded at the end of the circuit. Block B1 is deleted - the n+2 block is recorded at the end of the chain.

The system is scalable, the order of block formation by devices depends on the assigned id number. There is no possibility of random block creation. If a node that is outside of the queue offers to add a block to the chain, it is simply ignored. Also, each new block is checked for compliance with the remaining nodes, in order to prevent the substitution of information.

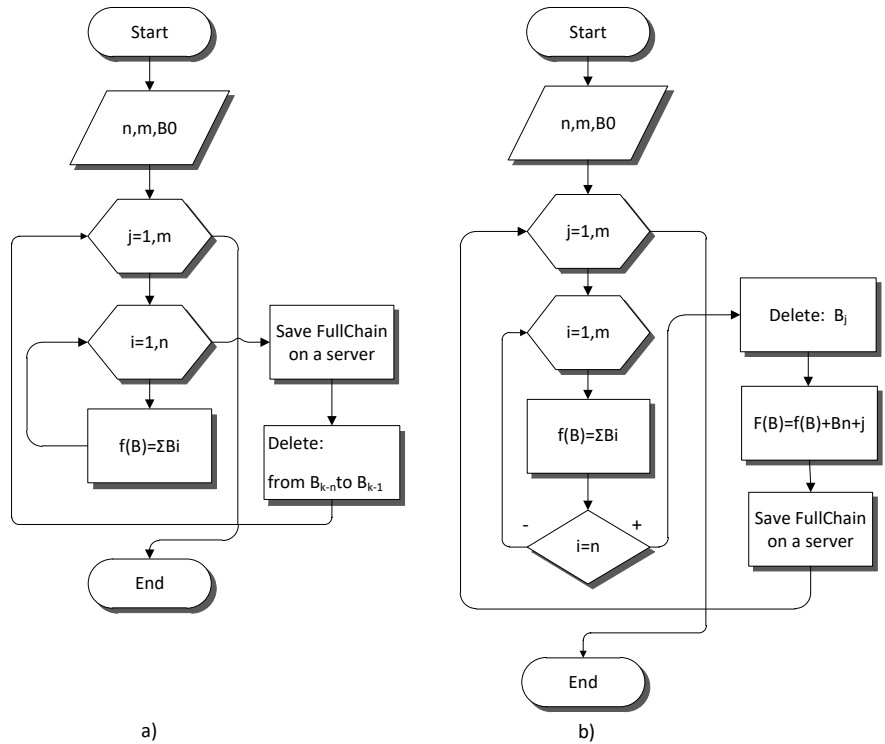

Fig. 6. The flowchart of block formation in a chain: a) when a predetermined number of blocks is removed; b) when blocks are removed step-by-step.

Given that this is a closed network, the identification of devices is done by comparing them with the list of authorized devices. The suggestion is that this list should be recorded on a separate, protected part of the flash memory of each node. The list is continuously updated in case of a disconnection or the replacement of the faulty device. The flowchart of BC formation in a network has the following form (Fig.6).

### 2.5 Segmented network. Problem formulation

To use the model proposed above, it is necessary that the number of WSN nodes be fixed at each particular time. This can be done by dividing the network to segments (Fig.7).

An example of such a network can be WSN of a smart city, where nodes 1-2-3-4 are stationary, and nodes c1-c12 can be, for example, smart cars. Thus, the centers of the subnets are stationary nodes. At each particular time, each subnet contains a fixed number of nodes. Also, each subnet builds its own part of the block, which is sent to the server. Nodes that are elements of several subnets retain the last block of the previous element of the chain for verifying the next element of the neighboring subnet. The server stores all the elements of the chain in a single blockchain. However, in this case there arises the problem of estimating the optimal number of elements of each subnet and the number of connections between its nodes for a given level of reliability. This will be discussed in the next section.

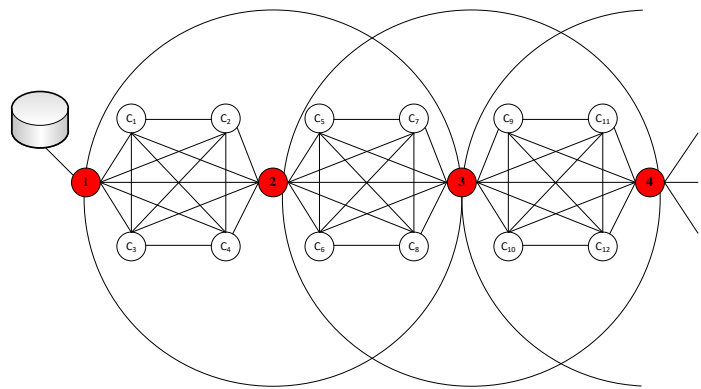

Fig. 7. The structure of WSN separation on a sub-segment when using mobile nodes.

### 2.5 Mathematical model

Mathematically, the structure of the complete chain is:

$$F = \bigcup_{i=0}^{\infty} B_i$$

Then, the part of the chain that will be removed from the memory of the node in each cycle will be as follows:

$$f_{del} = \bigcup_{i=0}^{n-1} B_i, n = const$$

Here $n$ - the number of nodes in the local network.

The general function will have the form of a matrix consisting of n columns and m rows.

For example, the matrix for 5 nodes, with end-to-end numbering of blocks, will have the following form for the m iterations:

Each of these is the sum of blocks in a chain for one cycle without taking into account the genesis block B0.

$$F = \begin{pmatrix} B_{11} & B_{12} & ... & B_{1(n-1)} & B_{1n} \\ B_{21} & B_{22} & ... & B_{2(n-1)} & B_{2n} \\ \vdots & \vdots & \ddots & \vdots & \vdots \\ B_{m1} & B_{m2} & ... & B_{m(n-1)} & B_{mn} \end{pmatrix}. \qquad (1)$$

Then, the analytic expression, which describes the structure of the complete chain, will have the form of (2).

$$F = B_0 \bigcup_{i,j \in S} B_{i,j}, S = [i = 1 \ldots n, j = 1 \ldots m]. \quad (2)$$

where the notation is as follows: j - an amount of elements in each row of matrix (1); n - number of the elements in the row; i – number of rows. Also, it should be borne in mind that the elements $B_{in} = B_{(i+1)1}$.

## 3  MAIN RESULTS

In this section, we consider the implementation of the proposed method for Blockchain formation.

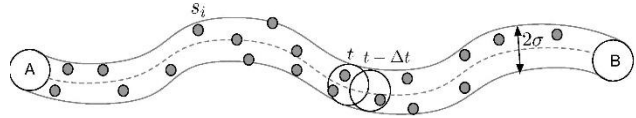

Fig. 8. The principle of building Blockchain for WSN, which consists of sub-segments, for mobile nodes.

Our procedure works as follows. First we create a linear arrangement of sensors as shown in Fig.8.

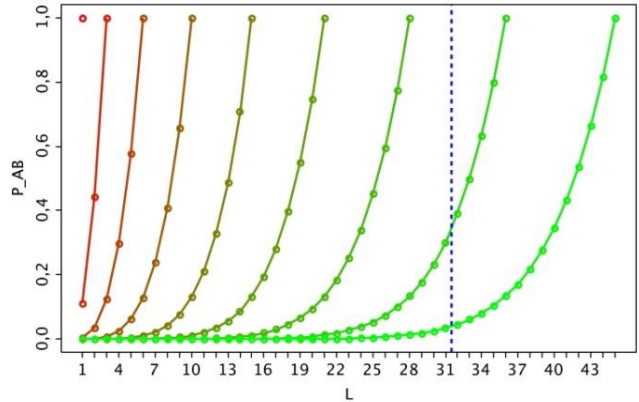

Fig. 9. Probability of finding a connected path between A and B as a function of the number of nodes "n" and the number of edges "L". The dashed blue line represents the limit 0.7Lmax for n=10 (and Lmax=45).

In this Fig. 9 we plot the probability of finding a connected path between A and B as a function of the number of nodes "n" and the number of edges "L". It is computed using a random graph model where the connection probability between two randomly chosen nodes is "p" = Lmax/L, where Lmax=n(n-1)/2. Each curve ranges L from 1 to Lmax for each n. The dashed blue line represents the limit 0.7Lmax for n=10 (and Lmax=45) which is the minimum connectivity threshold found in [19].

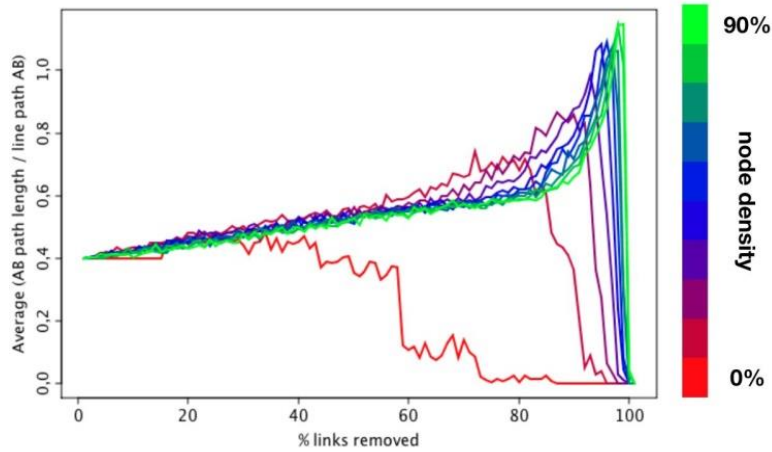

Fig. 10. Probability of finding a connected path between A and B as a function of the number of nodes "n" and the number of edges "L". The dashed blue line represents the limit 0.7Lmax for n=10 (and Lmax=45).

These represent fixed sensors along the way. Then we augment the sensor set by randomly spreading additional sensor over the area so that we reach a target sensor density. Each fixed sensor can transmit/receive signals within a radius (see circles in Fig.8). Therefore we create a network consisting of the union of the complete graphs of sensors lying inside each line sensor radius. Then we randomly remove links and check if a path between the start and the end line nodes can still be built. We did Monte Carlo tests to numerically find the probability of finding a path when a portion of the links were randomly removed for different node densities. In Fig.10 we show the results of this analysis. Here we analyse the length of alternative shortest paths compared to the length AB paths along the horizontal line.

As it can be noticed, when we increase the level of attack (proportion of links removed) the network is resilient to provide alternative paths until its break down. At this point no alternative paths are possible. As expected, in sparse scenarios (low sensor density) this break down occurs at smaller attack intensities.

## 4 SUMMARY AND DISCUSSION

The findings outlined in this article can be applied to at least two fields: Wireless Sensor Networks and the Internet of Things. Clearly, this contribution is just a first step in the understanding of short and partially connected BC.

The simulation results showed that with increasing attack density (increasing the number of lost connections and nodes) the network remains stable and the Blockchain can be built. It should be noted that some of the blocks (information from blocked nodes) can be lost. The number of lost blocks depends on the density of the sensors and the intensity of the attack. It should be taken into account that the reliability of such a network depends on the number of nodes at each moment of time in each separate sub-segment. Also, the minimum value of nodes is found, which should participate in the construction of the chain, in order to avoid network interruption.

However, the results of the work clearly show the possibility of constructing "a Rolling Blockchain", using mobile nodes when the the start and the end of the route are given. The problem still needs further elaboration in order to foster more robust implementations. For instance, we neglected the issues of security analysis and protection against hacking of the proposed method. In addition, the issue of having to use the Merkle tree for this type of network and chain has been left open. In future works we will research other topological models and how the use of the Merkle tree in the proposed algorithm, will affect the resource of the node batteries and what the ratio - increasing the stability/power consumption of the node is profitable for using in long Blockchchains. We will also consider the problem of calculating the optimal size of the memory used, on the basis of the elements available in the network, in order to optimize their performance.

## 5 Acknowledgments

This research was partially supported by the Regional government of Provincia Autonoma di Trento (Italy) and Bruno Kessler Foundation (Italy) under the Secure Blockchain-based Application project.

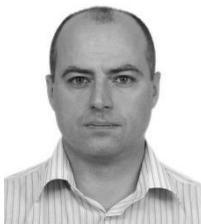

**Sergii Kushch** has got Bachelor's Degree in Radio Engineering at the Electronic Technologies Faculty of Cherkasy Engineering and Technological Institute in 1998. In 1999 he has got a diploma of Radio Engineer. The area of his research was statistical Radio Engineering. In 2009 Sergii Kushch has got Master's Degree in Radio Engineering at the Faculty of Electronic Technologies of Cherkasy State Technological University. The title of his diploma thesis was "Synthesis of statistical pattern recognition algorithms of signals that have been taken against a background of non-Gaussian noise". In 2009 he was invited to PhD school which he success finished in 2012. The title of his thesis was "Methods of multivariate realization the components of computer systems". Since 2012 he worked at Cherkasy State Technological University on the positions of lecturer, Sr. lecturer and Associate professor of Department of Computer Science and Information Security. He published a number of scientific papers in international and national journals and took part in international and national conferences. Currently, he works as researcher in Foundation Bruno Kessler, Italy in Security and Trust research unit with modification Blockchain technology for an increasing protection personal information EU citizens.

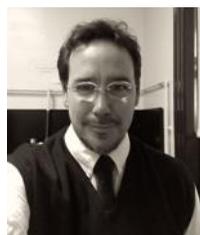

**Francisco Prieto Castrillo** With a background in theoretical physics, Francisco Prieto Castrillo has worked in many, seemingly distant fields; statistical physics, artificial intelligence, smart energy networks and seismic engineering. The backbone of his involvement in all these activities is his enthusiasm for the analysis of the self-organization phenomena in complex systems. This includes the way a structure resists an earthquake or the patterns of collective social behaviour.

Francisco Prieto Castrillo is a researcher in the BISITE group at University of Salamanca. His former position was as postdoctoral researcher at the New England Complex Systems Institute (NECSI), a centre which has been instrumental in the development of complex adaptive systems. Currently he is affiliated to MIT and to Harvard University. At MIT he collaborates with the Alex "Sandy" Pentland's Human Dynamics Laboratory in the MIT- MediaLab. At the Harvard T. H. Chan School of Public Health, he analyses different problems related to social sciences using data mining techniques, machine learning and complex systems.